\pdfoutput=1
\documentclass[twocolumn,showpacs,preprintnumbers,amsmath,amssymb, prl]{revtex4}

\usepackage{dcolumn}
\usepackage{bm}
\usepackage{hyperref} 
\usepackage{amsmath}
\usepackage{amsfonts}     
\usepackage{graphics}  
\usepackage{psfrag} 
\usepackage{graphicx}    
\usepackage{epsfig}    
\usepackage{rotating}  
 \usepackage[latin1]{inputenc}
 \usepackage{color}
 \usepackage{rotating}
 \usepackage{color}

\definecolor{DarkGreen}{rgb}{0,0.5,0}
\definecolor{Grey}{rgb}{0.5,0.5,0.5}
\definecolor{DarkYellow}{rgb}{1,0.7,0}
\definecolor{Violet}{rgb}{0.6,0.0,0.7}
\definecolor{Brown}{rgb}{0.5,0.3,0}

\begin{document}


\title{Entropy production of cyclic population dynamics}
\author{Benjamin Andrae$^{1}$}
\author{Jonas Cremer$^{1}$}
\author{Tobias Reichenbach$^{2}$}
\author{Erwin Frey$^{1}$}

\affiliation{$^{1}$Arnold Sommerfeld Center for Theoretical Physics (ASC) and Center for NanoScience (CeNS), LMU M\"unchen, Theresienstra{\ss}e 37, 80333 M\"unchen, Germany\\
$^{2}$Howard Hughes Medical Institute and Laboratory of Sensory Neuroscience, The Rockefeller University, 1230 York Avenue, New York, NY 10065-6399}


\date{\today}
\begin{abstract}
Entropy serves as a central observable in equilibrium thermodynamics.  However, many biological and ecological systems operate far from thermal equilibrium. Here we show that entropy production can characterize the  behavior of such nonequilibrium systems. To this end we calculate the entropy production for a population model that displays nonequilibrium behavior resulting from cyclic competition. At a critical point the dynamics exhibits a transition from large, limit-cycle like oscillations to small, erratic oscillations. We show that the entropy production peaks very close to the  critical point and tends to zero upon deviating from it. We further provide analytical methods for computing the entropy production which agree excellently with numerical simulations.
\end{abstract}

\maketitle

The study of complex  systems with a large number of interacting particles requests global observables that  characterize their behavior. Modern statistical mechanics has successfully identified, interpreted and applied such observables for  equilibrium systems. One of these observables is the entropy which allows for predictions of a system's behavior through the second law of thermodynamics -- an isolated system's entropy cannot decrease.
Identifying similar principles for  non-equilibrium systems, however,  proves elusive. Neither a characteristic global observable nor a universal principle have been identified in a general way.  While also in nonequilibrium the entropy production has been proposed as a useful observable~\cite{qiang,schlogl}, and different principles governing its behavior have been suggested~\cite{Jaynes, Prigogine}  problems  arise from different  employed definitions of  entropy  and approaches to nonequilibrium dynamics~\cite{schlogl,goldstein-2004-193,Seifert}.

In this Letter we demonstrate that entropy production can successfully characterize  ecological systems  with cyclic competition.
Ecological systems  display  a wide variety of nonlinear and  nonequilibrium behavior.  Random interactions between individuals and the finiteness of the population lead to intrinsic stochasticity. Nonequilibrium results when interactions between individuals of different species include cyclic dependencies where a species $A_1$ benefits from and suppresses a species  $A_2$, while $A_2$ benefits from and suppresses a species $A_3$ and so on, with some species $A_k$ of the resulting chain benefitting from and suppressing species $A_1$. Such cycles can lead to erratic or limit-cycle oscillations in the steady state of the population dynamics~\cite{may-1975-29,frachebourg-1996-54,mckane-2005-94,traulsen-2006-74,szabo-2007-446,TRaMMaEF07,claussen-2008-100,cremer-2008-63,boland-2009-79,galla-2009-103}. Experimental observations of cyclic dynamics and corresponding  oscillations have, amongst others, been documented for mating behavior of lizards in costal California~\cite{sinervo-1996-340} and  in microbial laboratory communities~\cite{kerr-2002-418}.

The dynamics of ecological systems can be conveniently described as a Markovian stochastic process  through a master equation,
\begin{equation}
\partial_t P_i(t)=\sum_j \left[ \omega_i^j P_j(t) - \omega_j^iP_i(t)  \right]\,,
\label{eq:master}
\end{equation}
in which $P_i(t)$ denotes the probability of finding the system in a certain state $i$ at time $t$ and $\omega_i^j$ is the transition probability from state $j$ to state $i$.
The associated mean entropy production $\dot{S}$ of the system follows as
\begin{equation}
\dot{S} = \frac{1}{2}\sum_{i,j} \left[\omega _i^j P_j(t) - \omega _j^i P_i(t)\right] \ln\left [\frac{\omega_i^j P_j(t)}{\omega_j^i P_i(t)}\right ]\label{Stot}.
\end{equation}
Equation~(\ref{Stot}) can be  obtained through considering the difference between forward and backward entropy per unit time of the stochastic process, Equation~(\ref{eq:master})~\cite{Gaspard}.
Equation~(\ref{Stot}) follows also 
 as the temporal derivative of the system's Gibbs entropy together with  a term describing the total increase of thermodynamic entropy in the reservoirs to which the system is coupled~\cite{Schnakenberg}.  
For steady states defined by $\partial_t P_i(t)=0$, as we consider in this Letter, the entropy production simplifies to
\begin{equation}
\dot{S} = \frac{1}{2}\sum_{i,j} \left[\omega _i^j P_j - \omega _j^i P_i\right] \ln\left [\frac{\omega_i^j }{\omega_j^i}\right ]\,.
\label{eq:ent_steady}
\end{equation} 

It follows from Equation~(\ref{Stot}) that the entropy production vanishes  if and only if the system  obeys detailed balance, $\omega _i^j P_j = \omega _j^i P_i$. Indeed, detailed balance represents the notion of   thermodynamic equilibrium in the framework of the master equation. Cyclic population dynamics violates detailed balance; the computation and discussion of the associated entropy production is the scope of this Letter.

Consider a simple model for cyclic population dynamics of three species $A,B,$ and $C$. Interactions are formulated as chemical reactions:
\begin{eqnarray}
AB \overset{k}{\longrightarrow} AA\,, \hspace*{0.5cm} A \overset{m}{\longleftrightarrow} B\,, \nonumber \\
BC \overset{k}{\longrightarrow} BB\,, \hspace*{0.5cm} B \overset{m}{\longleftrightarrow} C\,, \nonumber \\
CA \overset{k}{\longrightarrow} CC\,, \hspace*{0.5cm} C \overset{m}{\longleftrightarrow} A\,.\label{model}
\label{eq:react}
\end{eqnarray}
The reactions on the left describe cyclic competition: $A$ outperforms $B$ but is beaten by $C$, and $C$ is taken over by $B$ in turn. The reactions on the right correspond to spontaneous mutations between the three species.

\begin{figure}
\centering
\begin{tabular}{ll}
(a)\hspace*{0.9cm} $m<m_c$ &(b)\hspace*{0.9cm}  $m>m_c$\\
\includegraphics[height=3.7cm]{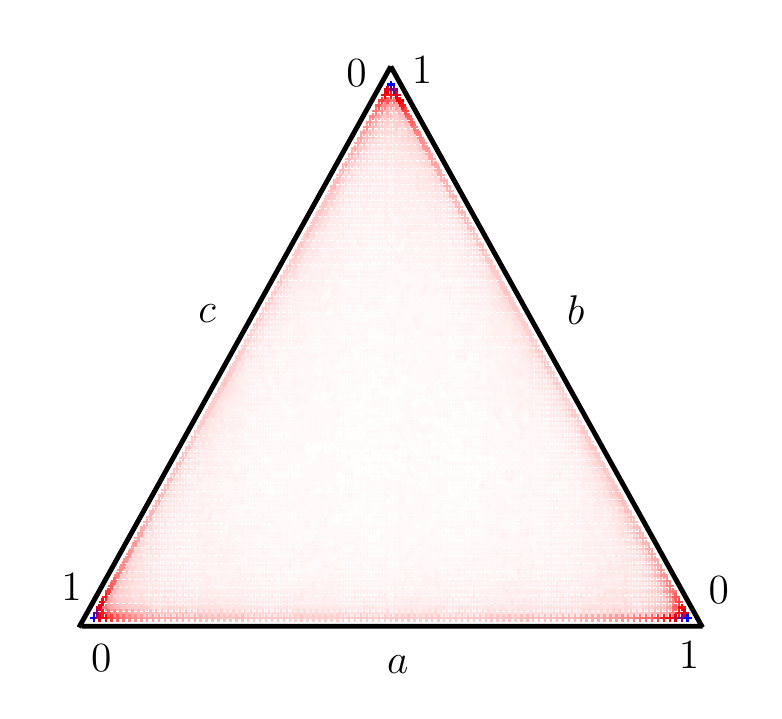}&
\includegraphics[height=3.7cm]{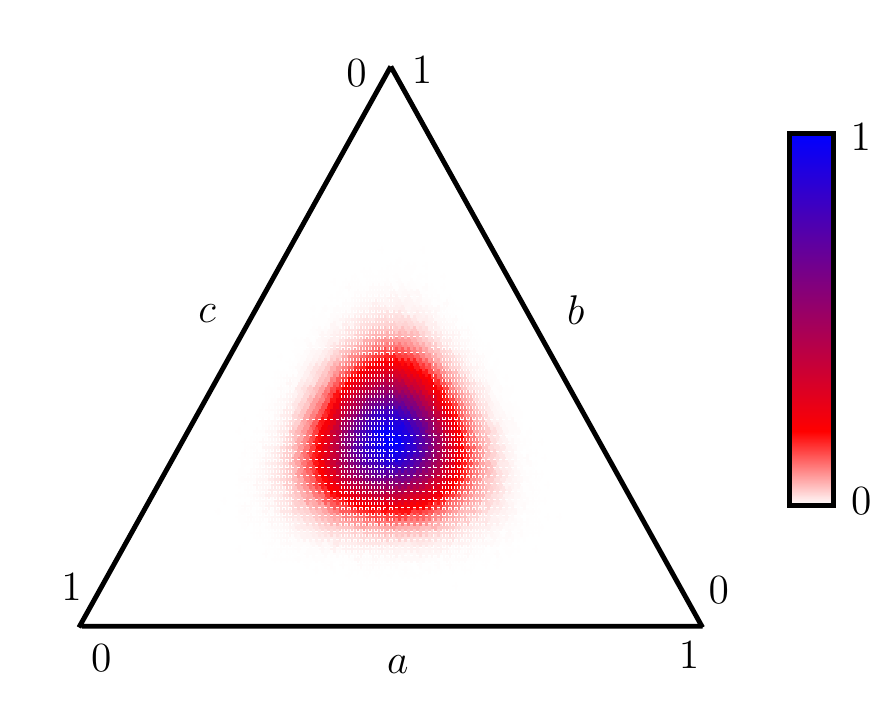}
\end{tabular}
\caption{(color online) Probability distributions  ($k=1,~N=100$).  (a) For a mutation rate  $m=0.003$ smaller than $m_c=k/(2N)$ the probability distribution is concentrated near the edges and particularly near the corners of the phase space. (b) A  mutation rate $m=0.1$ larger than $m_c$  leads to a gaussian distribution around the center. }\label{fig:states}
\end{figure}
The  population model defined by the reactions~(\ref{eq:react}) exhibits a critical mutation rate that, in the resulting non-equilibrium steady state, delineates large oscillations in the species densities from only small ones. Let us introduce this critical mutation rate first. The reactions~(\ref{eq:react}) conserve the total number $N$ of interacting individuals. The densities $a,b$, and $c$ of species $A,B,C$ therefore sum up to one: $a+b+c=1$ and the population's state space is the simplex $S_3$, see Fig.~\ref{fig:states}. Numerical simulations indicate that small values of the mutation rate $m$ lead to large oscillations between the densities of the three species; the probability distribution is highest close to the corners of the simplex [Fig.~\ref{fig:states}(a)]. Large values of $m$, on the contrary, lead to an approximately gaussian probability distribution around the simplex center [Fig.~\ref{fig:states}(b)]. Erratic oscillations occur at small amplitudes~\cite{mckane-2005-94}. 

The system's behavior can be analytically described by an approximate Fokker-Planck equation. A systematic expansion in the system size $N$ yields an equation for the temporal evolution of the probability distribution $P({\bm s},t)$ of the densities ${\bm s}=(a,b)$ at time $t$:
\begin{align}
\partial_t P({\bm s},t) = -  \partial_i[ \alpha_i({\bm s}) P({\bm s},t)]
 + \frac{1}{2} \partial_i \partial_j [\beta_{ij}({\bm s})P({\bm s},t)]\,,\label{eq:fokk}
\end{align}
in which the indices $i,j$ run from $1$ to $2$; the summation convention implies summation over them. The density $c$ follows as $c=1-a-b$.
The coefficients read 
\begin{align}
\alpha_i({\bm s})=& \left[m(1-3s_i) + ks_i(s_{i+1}-s_{i+2})\right]\,, \cr
\beta_{ii}({\bm s})=&N^{-1}\left[m(1+s_i) + ks_i(s_{i+1}+s_{i+2})\right]\,,\cr
\beta_{ij}({\bm s})=&-N^{-1}\left[m(s_i+s_j) +ks_is_j\right]\quad \text{for}\quad i\neq j\,,
\label{coeff}
\end{align}
where the indices are understood as modulus $3$ and $s_3=c$.
The terms containing $\alpha$ describe the deterministic part of the temporal evolution. In the absence of fluctuations, the reactions for cyclic dominance lead to neutrally stable oscillations around the internal fixed point ${\bm s}_*=(1/3,1/3,1/3)$, while the spontaneous mutations render the internal fixed point stable. Demographic fluctuations are, for large system sizes $N$, inversely proportional to $\sqrt{N}$ and enter the Fokker-Planck equation~(\ref{eq:fokk}) through the terms containing $\beta$. They induce a stochastic drift away from the internal fixed point towards the boundaries of the phase space. The Fokker-Planck equation~(\ref{eq:fokk}) shows that the competition between the deterministic and the stochastic effects leads, at a  critical mutation rate $m_c=k/(2N)$, to a uniform probability distribution. Certain deviations from the uniform distribution occur near the phase space boundaries where the discreteness of the phase space becomes relevant and the continuous formulation through the Fokker-Planck equation does not hold. For small mutation rate, $m<m_c$, fluctuations dominate and drive the system towards the boundary. In the absence of mutations the corner states are absorbing and the system goes extinct~\cite{berr-2007-102}. An arbitrary small mutation rate, however, leads to sustained species coexistence and oscillations. In the opposite case, when $m>m_c$, the deterministic dynamics centers the probability distribution around the internal fixed point.

The cyclic population dynamics yields a non-equilibrium steady state that is characterized by oscillations, large or small, around the internal fixed point. What is the resulting entropy production and how does it relate to the regimes of small, critical and large mutation rates outlined above?

\begin{figure}
\centering
\includegraphics[width=7.2cm]{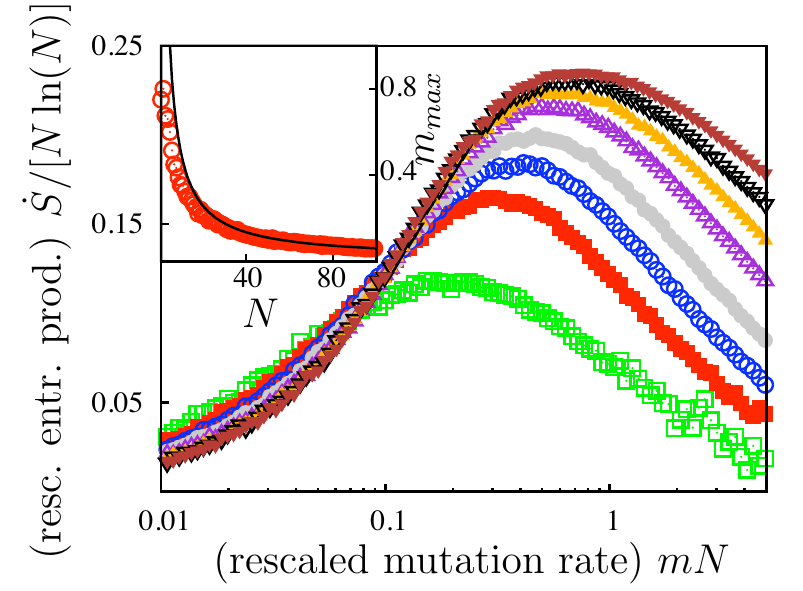}
\caption{(color online) Entropy production in the steady state for different system sizes
(\textcolor{green}{\scriptsize  $\Box$}, $N=2$; \textcolor{red}{$\blacksquare$}, $N=5$; \textcolor{blue}{\large $\circ$}, $N=10$; \textcolor{Grey}{\large $\bullet$}, $N=20$; \textcolor{Violet}{\scriptsize  $\bigtriangleup$}, $N=50$; \textcolor{DarkYellow}{$\blacktriangle$}, $N=100$; \textcolor{black}{\scriptsize  $\bigtriangledown$}, $N=200$; \textcolor{Brown}{$\blacktriangledown$}, $N=400$). 
The entropy production vanishes for very high and very low mutation rates and exhibits a maximum at an intermediate value $m_\text{max}$. The value $m_\text{max}$ is near the critical mutation rate as shown in the inset where the black line indicates  $m_c+0.001$, and red circles represent data obtained from simulations. }\label{fig:entropy}
\end{figure}
To tackle this question we have carried out extensive numerical simulations of the stochastic system employing the Gillespie algorithm~\cite{gillespie-1977-81}. Throughout our simulations we have considered $k=1$  which defines the time-scale. 
Numerical results from computer simulations of the stochastic system show that the entropy production peaks at a certain value $m_\text{max}$ of the the mutation rate (Fig.~\ref{fig:entropy}). The value $m_\text{max}$ approximately equals the critical mutation rate, $m_\text{max}\approx m_c$ (Fig.~\ref{fig:entropy} inset). Small deviations from this behavior arise for the probability distribution at the critical mutation rate is not uniform near the boundaries as mentioned above. 

Analytical understanding of the entropy production in the regimes of small, critical, and large mutation rates is feasible through the Fokker-Planck equation~(\ref{eq:fokk}). To this end we employ a continuous version of the entropy production~(\ref{eq:ent_steady}),
\begin{equation}
\dot{S} = \frac{1}{2}\int d{\bm r}\int d{\bm s} \left[\omega _{\bm s}^{\bm r} P({\bm s}) -\omega _{\bm r}^{\bm s} P({\bm r})\right] \ln\left (\frac{\omega _{\bm s}^{\bm r}}{\omega _{\bm r}^{\bm s}}\right )\,,
\label{eq:ent_steady_cont}
\end{equation} 
where integration is over all states ${\bm r,\bm s}$ of the phase space.

The entropy production in form of Eq.~(\ref{eq:ent_steady_cont}) can readily be evaluated at the critical mutation rate $m_c$. The probability distribution is uniform according to the Fokker-Planck equation~(\ref{eq:fokk}); we obtain
\begin{equation}
\dot{S}_{m=m_c} = \frac{3}{144}kN\left[12 \ln(N) -13 + 6\ln(4) \right].\label{rescale}
\end{equation}
For moderate and large $N$  the term $N\ln(N)$ on the right-hand side dominates the entropy production. Stochastic simulations confirm this behavior (Fig.~\ref{fig:entropy}).

In the regime of large mutation rates,  $m>m_c$, we need to calculate the probability density in the steady state to compute the entropy production. We obtain the probability density by using polar coordinates $(r,\phi)$ centered at  the internal fixed point. We then simplify the Fokker-Planck equation~(\ref{eq:fokk}) through a van-Kampen approximation  for the coefficients (\ref{coeff}): the latter are approximated by their values at the internal fixed point. The resulting Fokker-Planck equation is then solved by the gaussian distribution
\begin{equation}
P (r, \varphi) = \frac{1}{2\pi \sigma^2}\exp\left( \frac{-r^2}{2\sigma^2}\right)\,,
\label{eq:gauss}
\end{equation}
where $\sigma = \sqrt{\frac{k + 6m}{36 m N}}$.  The entropy production follows from Eq.~(\ref{eq:ent_steady_cont}) where the integral is evaluated by setting the upper boundary of the integral  to $\infty$ and an average over the angular dependence is taken:
\begin{equation}
\dot{S}_{m\gg m_c} = \frac{kN}{3} \ln\left(\frac{k}{3m} + 1\right).\label{Shighm}
\end{equation}
\begin{figure}
\centering
\begin{tabular}{ll}
(a) \hspace*{1.5cm} $m\ll m_c$ & (b)\hspace*{1.5cm} $m\gg m_c$ \\
\includegraphics[height=4.9cm]{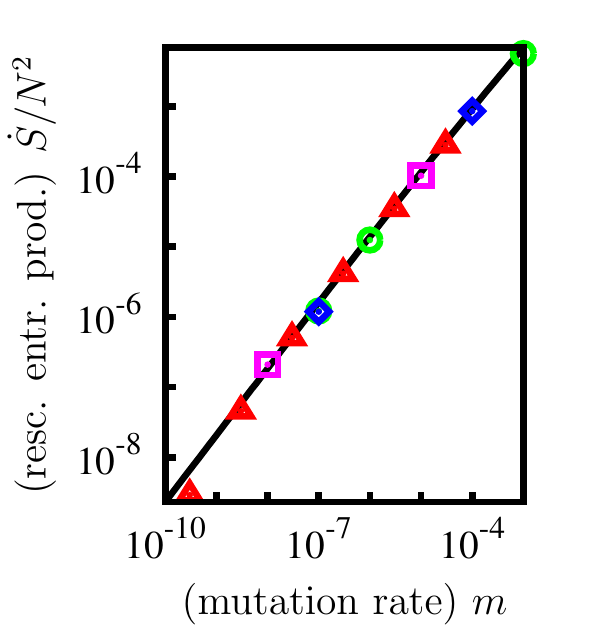}&
\includegraphics[height=4.9cm]{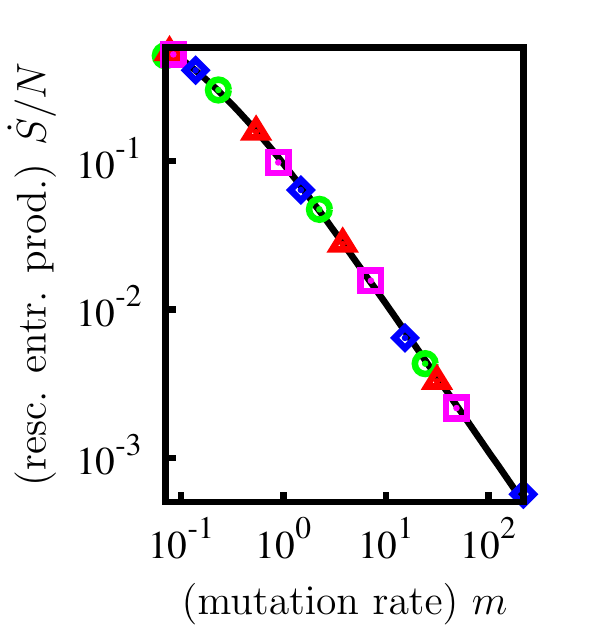}
\end{tabular}
\caption{(color online)  Entropy production in the limiting cases $m\ll m_c$ (a) and $m\gg m_c$ (b). Analytical results (black lines) agree excellently with simulations
(\textcolor{green}{\large $\circ$}, $N=30$; \textcolor{red}{\scriptsize  $\bigtriangleup$}, $N=100$; \textcolor{magenta}{\scriptsize  $\Box$}, $N=200$; \textcolor{blue}{\large$\diamond$}, $N=400$). 
The data confirm that the entropy production is proportional to the squared system size $N^2$ for small mutation rates and proportional to $N$ for large mutation rates. The simulation results further confirm that the entropy production decays as $m$ for $m\to 0$ and as $1/m$ for $m\to \infty$.}
\label{fig:large_small_m}
\end{figure}
This result agrees excellently with numerical simulations [Fig~\ref{fig:large_small_m}(b)].  The entropy  production~(\ref{Shighm})  depends linearly on the  system size $N$. This behavior arises because  the typical area in phase space explored by the dynamics is  proportional to $\sigma^2\sim 1/N$ and thus contains $N^2\sigma^2\sim N$ states. The continuity approximation employed in the Fokker-Planck equation~(\ref{eq:fokk})  holds for arbitrary large $m$,  since the width $\sigma$ of the probability distribution~ (\ref{eq:gauss}) remains finite as $m\rightarrow\infty$.

Expanding (\ref{Shighm}) for large values of  $m$ results in $\dot{S}_{m\gg m_c} = k^2N/(9m)$. The entropy production  vanishes as $m/k$ increases. Indeed, only the cyclic dynamics at rate $k$ underlies the nonequilibrium behavior and therefore entropy production, while the  mutations at rate $m$ obey detailed balance. 

When the mutation rate is small, $m\ll m_c$, the probability distribution is concentrated near the boundaries of the phase space [Fig.~\ref{fig:states} (a)]. The dynamics occurs predominantly along the boundary and can therefore be approximately described by considering only the boundary states. Because of the threefold symmetry it suffices to regard only one edge of the simplex with periodic boundary conditions. The concentration $x$ of one of the three species increases along this edge from $0$ to $1$ such that the cyclic dynamics drives the system to $x=1$. The deterministic part of the dynamics is given by 
\begin{equation}
\partial_t x=m(1-2x) - k(x-x^2)
\end{equation}
 and features a fixed point at $x_* = [(2m + k) - \sqrt{4m^2 + k^2}]/(2k)$. In the range of $m\ll m_c$, this fixed point is closer to $1$ than the  distance $1/N$ between two discrete states. We conclude that fluctuations will cause the system to exhibit a constant circular current in the steady state. The probability distribution $P(x)$ for $x \in [\frac{1}{N}, 1-\frac{1}{N}]$ can therefore be obtained as  solution to the Fokker-Planck equation $0 = -\partial_{x} \left\{[m(1-2x) - k(x-x^2)] P(x)\right\}$ where fluctuations have been ignored:
\begin{equation}
P(x) = \mathcal{N} \frac{1}{m-2mx-kx+kx^2},\label{borderapprox}
\end{equation}
with a normalization coefficient $\mathcal{N}$.
To determine the probability $P_0$ of a corner state, which turns out to be finite, fluctuations have to be included. $P_0$ can be obtained using the master equation and the values of $P(x=1/N)$ and $P(x=1-1/N)$. The normalization $\mathcal{N}$ follows from
\begin{equation}
3\left[  \int_{\frac{1}{N}}^{1 - \frac{1}{N}}P(x) dx+  \frac{2}{N}P_0  \right]=1\,.
\label{eq:N}
\end{equation}
 The factor $3$ arises because the phase space simplex possesses three edges.  For moderate and large system sizes $N$ we obtain $P_0=\mathcal{N}/(2m)$ which dominates the left-hand side of Equation~(\ref{eq:N}), such that $\mathcal{N}=Nm/3$. The resulting probability density can again be inserted into (\ref{eq:ent_steady_cont}) to provide an analytical result for the entropy production in the regime $m\ll \frac{k}{2N}\ll 1$:
\begin{equation}
\dot{S}_{m\ll m_c} = mN^2 \ln(m/k),\label{Slowm}
\end{equation}
 in perfect argeement with simulations [Fig~\ref{fig:large_small_m} (a)]. The entropy production for small mutation rates is proportional to the squared system size. Decreasing $m$ lowers the entropy production in proportion because mutations are the process that restart the cyclic dynamics once a corner state has been reached.  Mutations therefore limit the dynamics to a timescale proportional to $m$.

In conclusion, we have examined the global entropy production in the steady state of a cyclic population model. At a critical mutation rate  the system undergoes a transition from large oscillations along the phase space's boundary  to small erratic oscillations around an internal fixed point. The entropy production peaks very near the critical mutation rate and decreases to zero away from it.
We believe that, in a similar manner,  the entropy production can yield valuable information about the nonequilibrium steady state of other stochastic systems. Indeed, in a recently studied model, because of a non-fixed system size and the extensivity of the entropy production, the \emph{slope} of the entropy production peaks near a critical point~\cite{gaspard-2004-120}.  Within our approach of a fixed system size we have investigated a stochastic system that displays a supercritical Hopf bifurcation.  We found that the entropy production predicts the scale of the critical mutation rate: it peaks near the Hopf bifurcation,  at a mutation rate of about $1/4$ of the critical one~\cite{EPAPS}. Understanding the certain discrepancy between the maximum and the critical value may open a route to more general understanding of the role of entropy production.   Because of the  universality of the Hopf bifurcation we conclude that our approach is valid for a wide class of non-equilibrium systems, namely those that exhibit a transition from small, erratic oscillations to limit-cycle-like ones,  including systems with spatial degrees of freedom~\cite{szabo-2004-37}.

This research was supported by 
the German Excellence Initiative via the program 
`Nanosystems Initiative Munich' and the German Research Foundation via the SFB. 
TR12 `Symmetries and Universalities in Mesoscopic Systems'.
T. R. acknowledges support from the Alexander von Humboldt
Foundation through a fellowship.



\newpage

\onecolumngrid

\setcounter{page}{1}

\begin{center}
{ {\bf \Large
Entropy production of cyclic population dynamics} \\
\vspace*{0.5cm}
{ \large Benjamin Andrae, Jonas Cremer, Tobias Reichenbach,  and Erwin Frey}\\

\vspace*{0.5cm}
\bf \Large
Supplementary EPAPS Document: Entropy production and Hopf bifurcation
}
\end{center}
\vspace*{0.5cm}

Entropy production can characterize the behavior of a broad class of nonequilibrium systems. In this Supplementary Material we underpin this point through consideration of a nonequilibrium stochastic system that exhibitis a Hopf bifurcation. We show that the entropy production peaks in the vicinity of the bifurcation, where the behavior changes from noisy, erratic oscillations to larger limit-cycle oscillations. We conclude that the entropy production generally characterizes the behavior of  systems with limit cycles that fall into the universality class of the Hopf bifurcation. \\

Consider a stochastic system with species $A,B,C$ and empty sites $\oslash$ that obey the following reactions:
\begin{align}
AB&\stackrel{k}{\longrightarrow} A\oslash\,,\qquad  A\oslash \stackrel{l}{\longrightarrow}AA\,,\qquad \,A\stackrel{m}{\longleftrightarrow} B\,,\cr
BC&\stackrel{k}{\longrightarrow} B\oslash\,,\qquad  B\oslash \stackrel{l}{\longrightarrow}BB\,,\qquad B\stackrel{m}{\longleftrightarrow} C\,,\cr
CA&\stackrel{k}{\longrightarrow} C\oslash\,,\qquad  C\oslash \stackrel{l}{\longrightarrow}CC\,,\qquad \,C\stackrel{m}{\longleftrightarrow} A\,.
\label{eq:ml_react}
\end{align}  
The reactions with rates $k$ and $l$ can represent cyclic dominance of three species [11]. The corresponding deterministic rate equations have first  been proposed and analyzed by R.~M.~May and W.~J.~Leonard~[16].  The reactions with rate $m$ describe spontaneous mutations between the three species.

\begin{figure}[b]
\centering
\begin{tabular}{lll}
(a)\hspace*{1.6cm} $m<m_c$ & (b)\hspace*{1.6cm} $m=m_c$ & (c)\hspace*{1.6cm} $m>m_c$\\
\includegraphics[height=5cm]{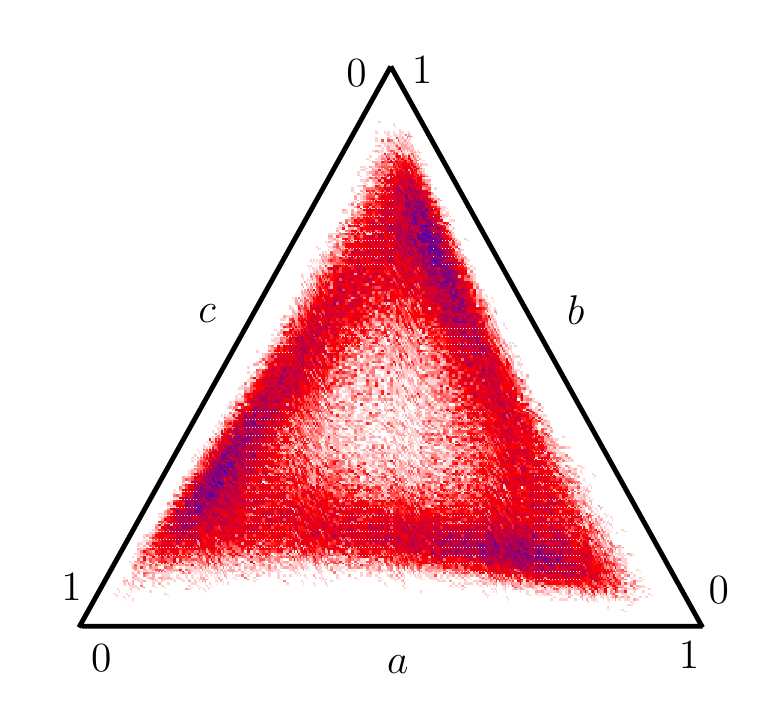}&
\includegraphics[height=5cm]{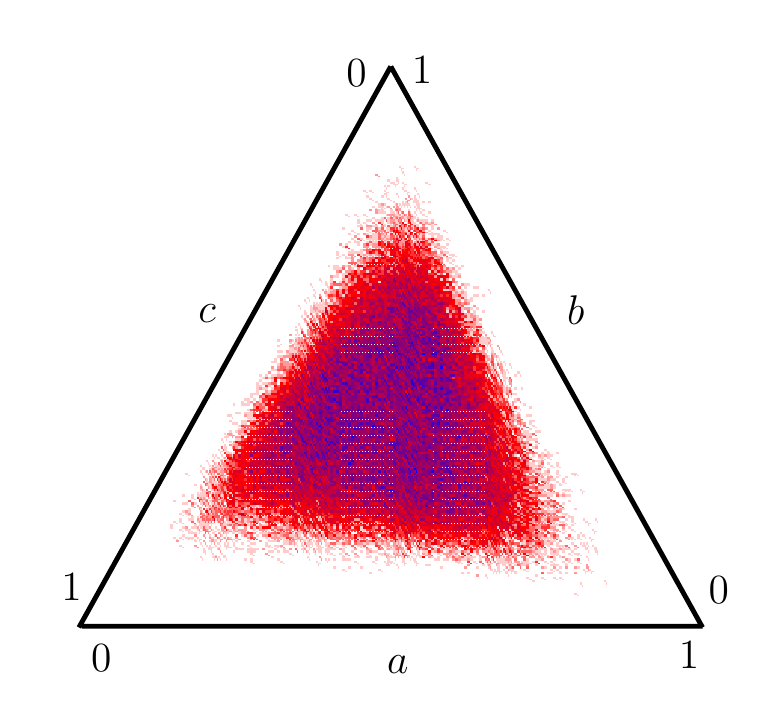}&
\includegraphics[height=5cm]{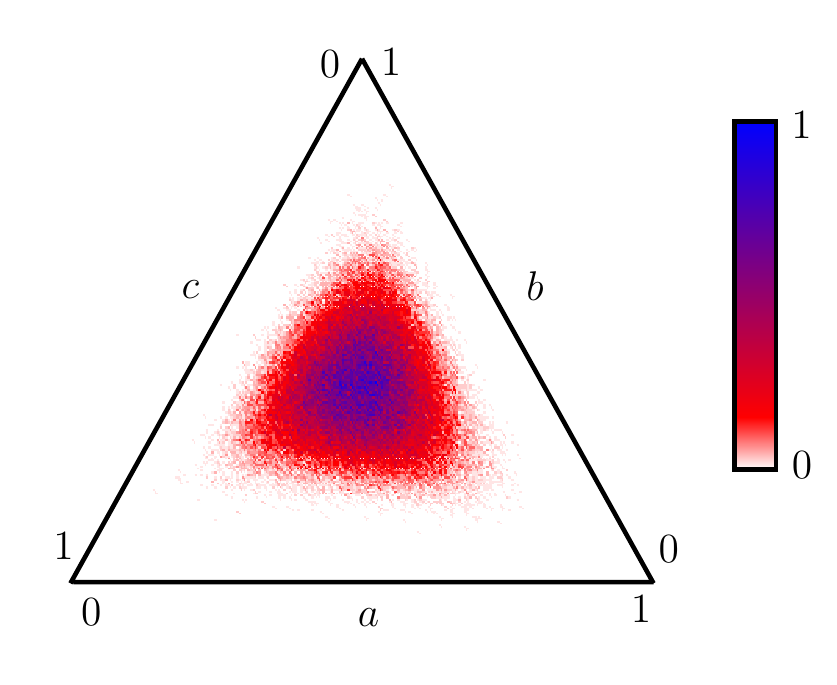}
\end{tabular}
\caption{(color online) Steady-state probability distributions for  $k=l=1,~N=100$, projected on the plane spanned by $(a,b,c)=(1,0,0),(0,1,0),(0,0,1)$.  (a) For a mutation rate  $m=0.022$ smaller than $m_c\approx0.042$ the probability distribution is concentrated along the limit cycle. (b) At the critical mutation rate $m_c$ a broad probability distribution centered around the reactive fixed point arises.  (c) A mutation rate $m=0.062$ larger than $m_c$  leads to a narrow, gaussian distribution around the reactive fixed point. }\label{fig:ml_states}
\end{figure}
The  deterministic equations for the temporal evolution of the concentrations $a,b,c$ of species $A,B,C$ follow from the reactions~(\ref{eq:ml_react}) as
\begin{eqnarray}
\partial_t a&=&a[l(1-\rho)-k c] + m(b+c-2a)\,,\cr
\partial_t b&=&b[l(1-\rho)-k a]+m(a+c-2b)\,,\cr
\partial_t c&=&c[l(1-\rho)-k b]+m(a+b-2c)\,.
\label{eq:ml_rate_eqs}
\end{eqnarray}
Linear stability analysis reveals the existence of a reactive fixed point at  $(a_*,b_*,c_*)=l/(3l+k)\cdot (1,1,1)$. This fixed point changes its stability at a critical mutation rate $m_c=kl/6/(3l+k)$ from a stable spiral point (above $m_c$) to an unstable spiral point (below $m_c$). Further analysis that takes the leading nonlinearities into account reveals that a supercritical Hopf bifurcation arises at $m_c$.  Above the critical mutation rate, for $m>m_c$, the stochastic system performs noisy erratic oscillations around the reactive fixed point. The steady-state probability distribution is approximately gaussian around the reactive fixed point, see Fig.~\ref{fig:ml_states}~(c) for a projection of the system's steady state onto the simplex spanned by the densities $a,b,c$.  Below the critical mutation rate, for $m<m_c$, a stable limit cycle forms. The stochastic dynamics leads to noisy trajectories along the limit cycle, see Fig.~\ref{fig:ml_states}~(a). At the critical mutation rate, as the linear terms in the deterministic equations vanish, a relatively broad, non-gaussian probability distribution centered at the reactive fixed point arises [Fig.~\ref{fig:ml_states}~(b)]. This behavior is similar to the one recently reported in Ref.~[13] where higher order nonlinearities render a spiral point stable while the linear terms vanish.

\begin{figure}[t]
\centering
\includegraphics[width=8.5cm]{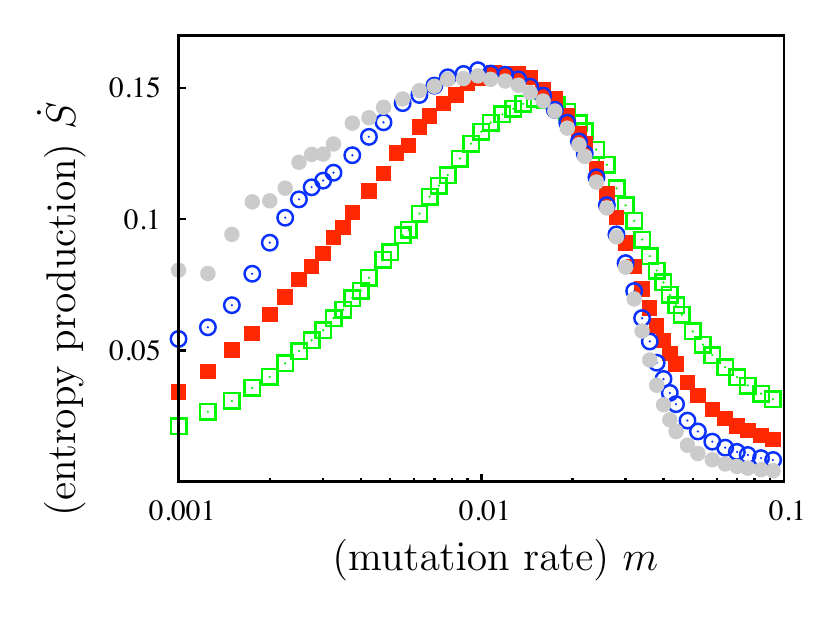}
\caption{(color online)  Entropy production in the steady state. Results  for different system sizes
(\textcolor{green}{\scriptsize  $\Box$}, $N=100$; \textcolor{red}{$\blacksquare$}, $N=200$; \textcolor{blue}{\large $\circ$}, $N=400$; \textcolor{Grey}{\large $\bullet$}, $N=800$) show that the entropy production peaks at a mutation rate $m_\text{max}\approx 0.01$ near the critical mutation rate $m_c\approx 0.042$. 
The entropy production vanishes both for smaller and higher mutation rates.}\label{fig:ml_entropy}
\end{figure}
We have performed extensive stochastic simulations of the stochastic system defined by the reactions~(\ref{eq:ml_react}). In these simulations we have left the rates $k,l$ constant at  $k=l=1$, defining the time-scale, and systematically varied the mutation rate $m$ as well as the system size $N$. In principle, a divergence in the entropy production can arise when the system reaches the boundary of the phase space. However, because the probability of these boundary states is exponentially suppressed, this effect can be ignored.

 For all considered system sizes the resulting entropy production  exhibits a maximum near the critical mutation rate $m_c$, see Fig.~\ref{fig:ml_entropy}. For system sizes above about $N=200$ the maximum of the entropy production arises at a value $m_\text{max}\approx 0.01$, about $1/4$ of the value of the critical mutation rate $m_c\approx 0.042$. For mutation rates much smaller and much larger than $m_\text{max}$ the entropy production tends to zero. The system's behavior therefore resembles the one reported in the main part of this Letter, underpinning the general usefulness of entropy production in characterizing nonequilibrium steady states. Understanding the certain discrepancy of $m_\text{max}$ and $m_c$ will yield further  insight into the relation between entropy production and critical nonequilibrium behavior.

\end{document}